\documentclass[final,times,onecolumn]{elsarticle}
\journal{Physica A}
\usepackage{subfig}
\usepackage{amsmath}
\usepackage{color}
\biboptions{sort&compress}

\begin{document}

\begin{frontmatter}
\title{Markov Chain Order estimation with Conditional Mutual Information}
\author{M. Papapetrou}\ead{mariapap@gen.auth.gr}
\author{D. Kugiumtzis}\ead{dkugiu@gen.auth.gr}
\address{Department of Mathematical, Physical and Computational Science, Faculty of Engineering, Aristotle University of Thessaloniki, 54124 Thessaloniki, Greece}

\begin{abstract}
We introduce the Conditional Mutual Information (CMI) for the estimation of the Markov chain order. For a Markov chain of $K$ symbols, we define CMI
of order $m$, $I_c(m)$, as the mutual information of two variables in the chain being $m$ time steps apart, conditioning on the intermediate
variables of the chain. We find approximate analytic significance limits based on the estimation bias of CMI and develop a randomization significance
test of $I_c(m)$, where the randomized symbol sequences are formed by random permutation of the components of the original symbol sequence. The
significance test is applied for increasing $m$ and the Markov chain order is estimated by the last order for which the null hypothesis is rejected.
We present the appropriateness of CMI-testing on Monte Carlo simulations and compare it to the Akaike and Bayesian information criteria, the maximal
fluctuation method (Peres-Shields estimator) and a likelihood ratio test for increasing orders using $\phi$-divergence. The order criterion of
CMI-testing turns out to be superior for orders larger than one, but its effectiveness for large orders depends on data availability. In view of the
results from the simulations, we interpret the estimated orders by the CMI-testing and the other criteria on genes and intergenic regions of DNA
chains.
\end{abstract}

\end{frontmatter}

\section{Introduction}
\label{intro}
\par
Let $\{x_t\}_{t=1}^N$ denote a symbol sequence generated by a Markov chain $\{X_t\}$, of an unknown order $L \geq 1$ in a discrete space of $K$
possible states $A=\{a_1,\ldots,a_K\}$. The objective is to estimate $L$ from the symbol sequence $\{x_t\}_{t=1}^N$ for a limited length $N$.

Many criteria for Markov chain order estimation have been proposed and evaluated in terms of their asymptotic properties. The Bayesian information
criterion (BIC) was proposed to render consistency of the popular Akaike information criterion (AIC) \cite{Tong75,Katz81,Guttorp95}. However, BIC was
found to perform worse than AIC for small sequence lengths, questioning the value of asymptotic properties in practical problems
\cite{Schwa78,Katz81,Csiszar00,Dalevi06}. A more recent and  general criterion than AIC and BIC is the efficient determination criterion (EDC),
opting for a penalty function from a wide range of possible such functions \cite{Zhao01}. Peres-Shields proposed in \cite{Pere05} the maximal
fluctuation method, which compares transition probabilities for words of increasing lengths, and Dalevi and Dubhashi \cite{Dale05} modified it for
practical settings and, instead of having to set a different threshold for each problem, they estimate the order from a sharp change in the
transition probabilities. They found that the Peres-Shields (PS) estimator is simpler, faster and more robust to noise than other criteria like AIC
and BIC \cite{Dale05}. Another method is that of global dependency level (GDL), also called relative entropy, using the $f$-divergence to measure the
discrepancy between two probability distributions \cite{Baig11}. GDL was found consistent and more efficient than AIC and BIC on relatively small
sequences. Finally, the method of Menendez et al \cite{Mene01,Mene06,Mene11} makes likelihood ratio tests for increasing orders using the
$\phi$-divergence measures \cite{Pardo06}. This procedure was found more powerful in tested cases than the existing chi-square and likelihood ratio
procedures, and it has also been applied to DNA \cite{Mene11}.

Here, we follow a different approach and estimate the Markov chain order from sequential hypothesis testing for the significance of the conditional
mutual information (CMI) for increasing orders $m$, denoted as $I_c(m)$. $I_c(m)$ is the mutual information of $x_i$ and $x_{i+m}$ conditioning on
the intermediate variables of the chain, $x_{i+1},\dotsc,x_{i+m-1}$. A significant $I_c(m)$ indicates that the order of the Markov chain is at least
$m$. Thus the repetition of the significance test of $I_c(m)$ for increasing $m$ allows for the estimation of the Markov chain order $L$ from the
last order $m$ for which the null hypothesis of zero CMI is rejected. We show that the significance bounds for $I_c(m)$ formed by means of
appropriate resampling are more accurate than the approximate analytic bounds we derived based on previous analytic results on the bias of entropy
\cite{Roul99}. We further compare the CMI testing with other criteria for order selection on simulated Markov chains and DNA sequences.

The structure of the paper is as follows. In Section~\ref{sec:CMI}, CMI is defined and estimated on symbol sequences, an analytic significance limit
of CMI is derived, and a randomization significance test is proposed, forming our method of CMI-testing for the estimation of the Markov chain order.
Other methods for estimating the Markov chain order are briefly presented. In Section~\ref{sec:Simulations}, we assess the
efficiency of the proposed CMI-testing and compare it to other order selection criteria on simulations of Markov chains produced by randomly chosen
transition probability matrices of different order, as well as transition probability matrices estimated on genes and intergenic regions of DNA
sequence. In Section~\ref{sec:DNA}, we apply the CMI testing to the two DNA sequences and investigate the limitations of order estimation in terms of
data size. Finally, concluding remarks are discussed in Section~\ref{sec:Discussion}.

\section{Conditional Mutual Information and Markov Chain Order Estimation}
\label{sec:CMI}

First we define CMI in terms of mutual information and subsequently entropies. The Shannon entropy expresses the information (or uncertainty) of a
random variable $Xt$
\[
H(X) = - \sum_{x} p(x)\ln{p(x)},
\]
where the sum is defined for all possible symbols (discrete values) $x \in A$, and $p(x)$ is the probability of $x$ occurring in the chain. The
definition of Shannon entropy is extended to a vector variable $\mathbf{X}_t=[X_t,X_{t-1},\ldots,X_{t-m+1}]$ from a stationary Markov chain
$\{X_t\}$, referred to as word of length $m$, and reads
\[
H(\mathbf{X}_t) = - \sum_{x_t,\ldots,x_{t-m+1}} p(\mathbf{x}_t)\ln p(\mathbf{x}_t),
\]
where $\mathbf{x}_t=\{x_t,x_{t-1},\ldots,x_{t-m+1}\} \in A^{m}$, $p(\mathbf{x}_t)$ is the probability of a word $\mathbf{x}_t$ occurring in the
chain, and the sum is over all possible words of $K$ symbols and length $m$.

The mutual information (MI) of two random variables in the Markov chain being $m$ time steps apart, denoted $I(m)=I(X_t;X_{t-m})$, is defined in
terms of entropy as \cite{Cover91}
\begin{equation}
I(m) = H(X_t) + H(X_{t-m}) - H(X_t,X_{t-m}) = \sum_{x_t,x_{t-m}} p(x_t,x_{t-m}) \ln{ \frac{p(x_t,x_{t-m})}{p(x_t)p(x_{t-m})}}. \label{eq:I(m)}
\end{equation}

While $I(1)$ quantifies the amount of information $X_{t-1}$ carries about $X_t$ and vice versa, $I(2)$ cannot be interpreted accordingly due to the
presence of $X_{t-1}$, and the information of $X_{t-2}$ about $X_t$, or part of it, may already be shared with $X_{t-1}$. Thus if we are after the
genuine information of $X_{t-2}$ about $X_t$, we need to account for the information of $X_{t-1}$ about $X_t$. This is indeed desired when we want to
estimate the memory of the process, i.e. the order of the Markov chain. The appropriate measure for this is the conditional mutual information (CMI).
CMI of order $m$ is defined as the mutual information of $X_t$ and $X_{t-m}$ conditioning on $X_{t-m+1},\ldots,X_{t-1}$ \cite{Cover91}
\begin{align}
I_c(m) & = I(X_t;X_{t-m}|X_{t-1},\ldots,X_{t-m+1}) \notag \\
& = I(X_t;X_{t-1},\ldots,X_{t-m}) - I(X_t;X_{t-1},\ldots,X_{t-m+1}) \notag \\
& = -H(X_t,\ldots,X_{t-m})+H(X_{t-1},\ldots,X_{t-m}) \notag \\ & + H(X_t,\ldots,X_{t-m+1}) -H(X_{t-1},\ldots,X_{t-m+1}) \notag  \\
& = \sum_{x_t,\ldots,x_{t-m}} p(x_t,\ldots,x_{t-m}) \ln{ \frac{p(x_t|x_{t-1},\ldots,x_{t-m})}{p(x_t|x_{t-1},\ldots,x_{t-m+1})}}. \label{eq:CMI}
\end{align}
CMI coincides with MI for successive random variables in the chain, that is $I_c(1)=I(1)$.

\subsection{Estimation of Conditional Mutual Information}
The estimation of CMI is given through the estimation of the joint probability and the conditional probabilities in (\ref{eq:CMI}) by the
corresponding relative frequencies. Specifically, the maximum likelihood estimate (MLE) of $p(x_t,x_{t-1},\ldots,x_{t-m+1})$ is
\[
\hat{p}(x_t,x_{t-1},\ldots,x_{t-m+1})=\frac{n_{i_1,\ldots,i_m}}{K^m},
\]
where $n_{i_1,\ldots,i_m}$ is the frequency of occurrence of a word $\{i_1,\ldots,i_m\}\in A^m$ in the symbol sequence $\{x_t\}_{t=1}^N$, defined as
$n_{i_1,\ldots,i_m}=\sum_{t=m}^N \mbox{I}(x_t=i_1,\ldots,x_{t-m+1}=i_m)$, where I denotes the indicator function. Respectively, the MLE of the
conditional probability $p(x_t|x_{t-1},\ldots,x_{t-m})$ is
\[
\hat{p}(x_t|x_{t-1},\ldots,x_{t-m}) = \frac{n_{i_1,\ldots,i_m,i_{m+1}}}{n_{i_1,\ldots,i_m}}.
\]
The estimate $\hat{I}_c(m)$ of $I_c(m)$, by substituting the probability estimates in (\ref{eq:CMI}), inherently suffers from the inefficiency of MLE
at high dimensions being less accurate with the increase of $m$ or $K$ and the decrease of $N$, but also more biased. It has been proved that entropy
estimation involves a negative bias, i.e. the estimated value is lower than the real one \cite{LiWe90,Roul99}. Consequently, MI estimation has
positive bias which increases with $K$ and $m$ \cite{LiWe90,Roul99}, and thus the estimation of CMI has also positive bias, as CMI is the difference
of two MI terms, where the arguments in the first MI term have jointly a dimension larger by one than that of the second MI term. Expressing the bias
of CMI as the difference of the bias of two MI terms, indicates that the CMI estimate has lower bias than the bias for the respective MI's, which has
been shown for continuous variables in \cite{Vlachos10}.

An approximate expression for the bias of the entropy estimate of a random variable of $K$ symbols from a sample of size $N$ is given by
\cite{Roul99}
\[
\hat{H}(X)-H(X) = -(K-1)/(2N).
\]
Noting that a word $\mathbf{X}_t$ of length $m$ is equivalent to a random variable $X$ in $K^m$ symbols, we can express in the same way the bias of
the entropy estimate for a word $\mathbf{X}_t$ from a symbol sequence of length $N$ as
\[
\hat{H}(\mathbf{X}_t)-H(\mathbf{X}_t) = -(K^m-1)/(2N).
\]
Substituting the expressions for the entropy bias in the definition of CMI in terms of entropies in (\ref{eq:CMI}), we find the following
approximation for the bias of the CMI estimate
\begin{equation}
\hat{I}_c(m) - I_c(m) = {K^{m-1}(K-1)^2}/{2N}. \label{eq:bias}
\end{equation}
Note that the approximate bias for $I(X_t;X_{t-m})$ for any $m$ is ${(K-1)^2}/{2N}$ derived from (\ref{eq:bias}) for $m=1$. From (\ref{eq:bias}) it
can be seen that the bias of CMI increases when any of $K$ and $m$ increases and $N$ decreases.

\subsection{Randomization test for the significance of CMI}
We use CMI to estimate the order $L$ of a Markov chain. The fundamental property of a Markov chain of order $L$ is
\[
p(X_t|X_{t-1},X_{t-2},\ldots,X_{t-L},X_{t-L-1},\ldots) = p(X_t|X_{t-1},X_{t-2},\ldots,X_{t-L}),
\]
meaning that the distribution of the variable $X_t$ of the Markov chain at time $t$ is determined in terms only of the preceding $L$ variables of the
chain. Thus for any lag order $m \le L$, we expect in general two variables $m$ time steps apart to be dependent given the $m-1$ intermediate
variables, and then $I_c(m) > 0$. On the other hand, for $m > L$ it must be $I_c(m) = 0$. Note that it is possible that $I_c(m) = 0$ for $m < L$, but
not for $m=L$, as then the Markov chain order would not be $L$. So, increasing the order $m$, we expect in general when $I_c(m) > 0$ and $I_c(m+1) =
0$ to have $m=L$. To account for complicated and rather unusual cases where $I_c(m+1) = 0$ occurs for $m+1<L$, we can extend the condition $I_c(m) >
0$ and $I_c(m+1) = 0$ to require also $I_c(m+2) = 0$, and even further up to some order $m+k$.

The condition $I_c(m+1) = 0$ for $m=L$ does not hold exactly when estimating CMI from finite symbol sequences, and we always have $\hat{I}_c(m+1)>0$
due to positive bias in the estimation of $I_c(m+1)$. To address this, a significance test of $I_c(m)$ for increasing $m$ has to be developed for the
null hypothesis $\mbox{H}_0: I_c(m)=0$. In the absence of a rigorous analytic null distribution of the test statistic $\hat{I}_c(m)$, we propose a
randomization test using an ensemble of resampled (actually randomized as we preserve the marginal distribution) symbol sequences in order to form
the empirical null distribution of $\hat{I}_c(m)$. The test is one-sided with alternative hypothesis $\mbox{H}_1: I_c(m)> 0$, as the estimation bias
of $I_c(m)$ is positive. The randomization test is developed in the following steps.
\begin{enumerate}
\item We generate $M$ randomized symbol sequences $\{x_t^{*1}\}_{t=1}^N,\ldots, \{x_t^{*M}\}_{t=1}^N$, by random permutation of the initial sequence
$\{x_t\}_{t=1}^N$. \item We compute $\hat{I}_c(m)$ on the original symbol sequence, denoted $\hat{I}_c^0(m)$, and on the $M$ randomized sequences,
denoted $\hat{I}_c^{*1}(m),\ldots,\hat{I}_c^{*M}(m)$. \item We reject $\mbox{H}_0$ if $\hat{I}_c^0(m)$ is at the right tail of the empirical null
distribution formed by $\hat{I}_c^{*1}(m),\ldots,\hat{I}_c^{*M}(m)$. To assess this we use rank ordering, where $r^0$ is the rank of $\hat{I}_c^0(m)$
in the ordered list of the $M+1$ values, assuming ascending order. The $p$-value of the one-sided test is $1 - (r^0-0.326)/(M+1+0.348)$ (this
correction for the empirical cumulative function is proposed in \cite{YuHu01}).
\end{enumerate}
The randomized sequences are by construction independent, but with the same marginal distribution as the original sequence, and therefore they are
consistent with $\mbox{H}_0$. The estimation of $L$ with the proposed CMI-testing involves sequential implementation of the randomization
significance test of $I_c(m)$ for increasing $m$, starting with $m=1$. The repetitive procedure stops at an order $m+1$ if no rejection of
$\mbox{H}_0$ is obtained and then $\hat{L}=m$. To avoid premature termination of the sequential testing, which however can only be expected in
special practical cases, the termination criterion may require that $\mbox{H}_0$ is not rejected for more than one orders exceeding $\hat{L}=m$. The
termination criterion in the CMI-testing does not require a maximum order to be defined, which constitutes a free parameter for other order selection
criteria \cite{Katz81,Dale05}.

\begin{figure}[htb]
\centerline{\hbox{\includegraphics[width=7cm]{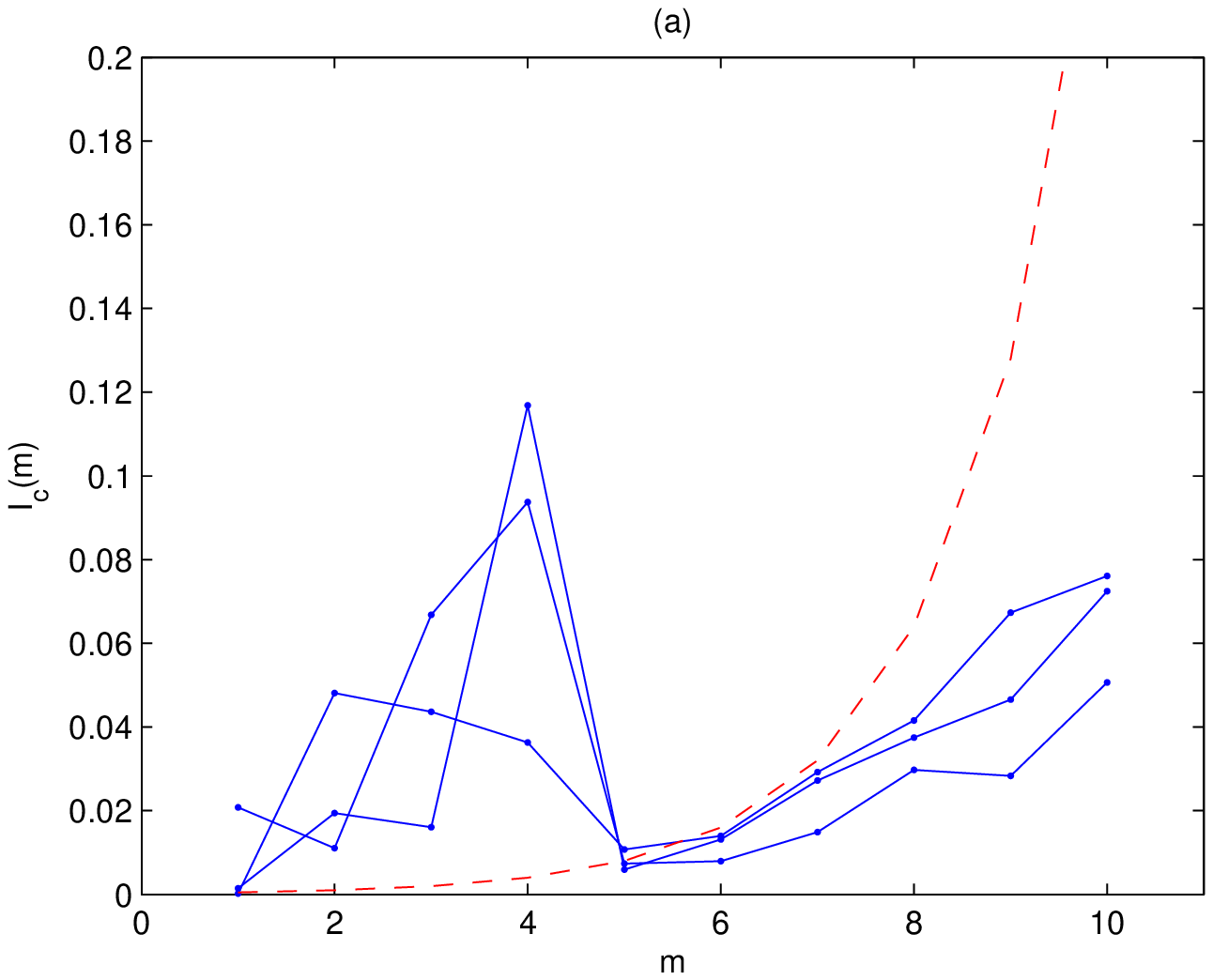}
\includegraphics[width=7cm]{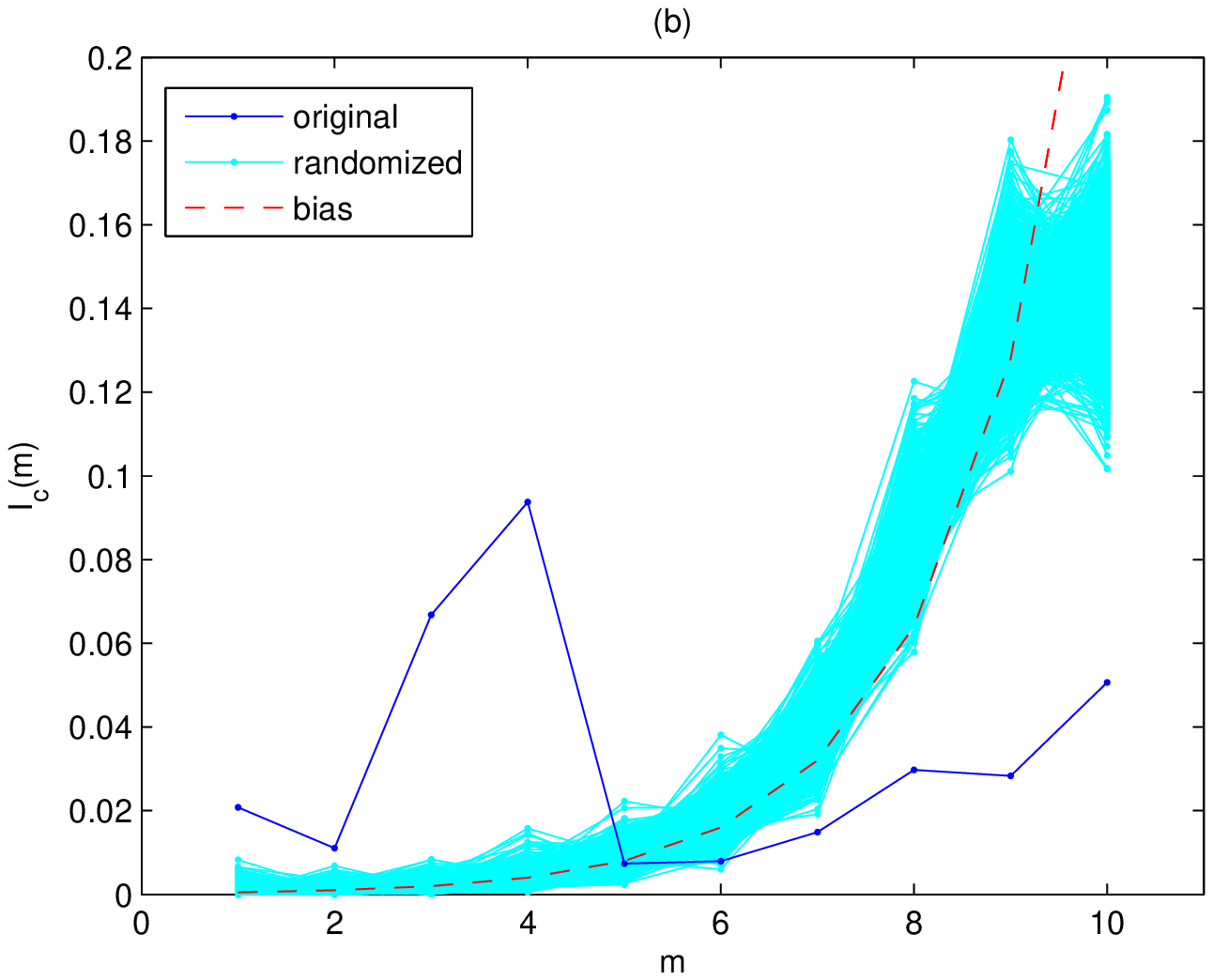}}}
\centerline{\includegraphics[width=7cm]{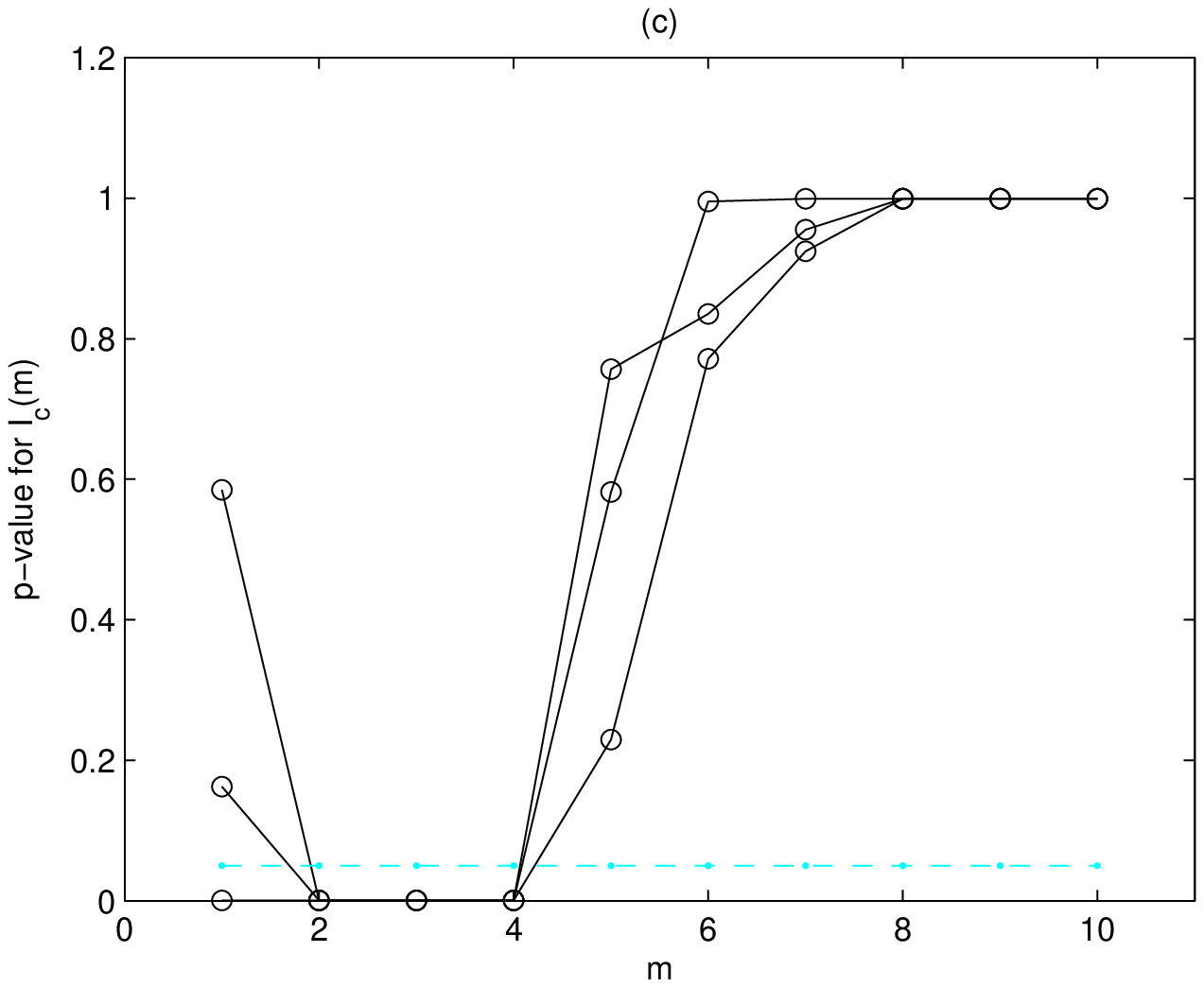}} \caption{(a) CMI vs order $m$ for three realizations of a Markov chain
determined by a randomly selected transition matrix ($K=2$, $L=4$, $N=1000$). Superimposed is the approximate bias estimate for zero CMI denoted with
a dashed grey (online red) line. (b) CMI vs order $m$ for one of the three realizations in (a) and for 1000 randomly shuffled sequences, as indicated
in the legend. (c) The $p$-value vs order of the randomization significance test for the three realizations in (a). The horizontal dashed grey
(online red) line is for the significance level $\alpha=0.05$.} 
\label{fig:example}
\end{figure}

We illustrate the proposed CMI-testing with an example of the estimation of the order $L=4$ of a Markov chain of $K=2$ symbols defined by a randomly
selected transition matrix. The CMI estimate $\hat{I}_c(m)$ for $m=1,\ldots,10$ computed on three symbol sequences of length $N=1000$ generated by
this Markov chain is shown in Figure~\ref{fig:example}a together with the approximate bias estimate for $I_c(m)=0$ derived in (\ref{eq:bias}).
For all three realizations, $\hat{I}_c(m)$ tends to increase for $m \le L$, then for $m=L+1$ drops at the level of the approximate bias, and further
increases for $m>L$ but does not exceed the approximate bias of zero CMI. Though the approximate bias seems to discriminate significant $I_c(m)$ for
$m\le L$ from insignificant $I_c(m)$ for $m>L$, it does not constitute an accurate upper bound of significance to be used as a criterion for the
estimation of $L$. Note that in Figure~\ref{fig:example}a not all three $\hat{I}_c(L+1)$ are below the approximate bias value. The use of randomized
sequences turns out to provide more accurate significance limits. As shown in Figure~\ref{fig:example}b for one of the three realizations, when $m
\le L$, $\hat{I}_c^0(m)$ is larger than any $\hat{I}_c^{*i}(m)$ for the $M=1000$ randomized sequences, but it is smaller or within the range of
$\hat{I}_c^{*i}(m)$, $i=1,\ldots,M$ when $m>L$. The $p$-values of the tests shown in Figure~\ref{fig:example}c for the three realizations confirms
the correct estimation of $L=4$ using the CMI-testing, being less than the significance limit $\alpha=0.05$ for $m \leq L$ and larger for $m
> L$. In the next Section, these findings are established by means of Monte Carlo simulations and compared to other known order estimation criteria.

\subsection{Other criteria for Markov chain order estimation}

There are various Markov chain order estimators in the literature
\citep{Baig11,Csiszar00,Dale05,Dalevi06,Guttorp95,Katz81,Mene01,Mene06,Mene11,Pardo06,Pere05,Schwa78,Tong75,Zhao01}, and we briefly discuss here the
most prominent ones that we also consider in the comparative study . The first is the well-known Akaike's information criterion (AIC)
\cite{Tong75,Katz81,Guttorp95}, which uses the Kullback-Leibler information to define the likelihood ratio (LR) statistic of $k$-th order versus
$L$-th order Markov chain
\[
n_{k,L}=-2\sum_{i=1}^{N}\ln\left(f(x_i|\hat{\theta}_k)/f(x_i|\hat{\theta}_L)\right),
\]
where $\hat{\theta}_L$ is the unrestricted maximum likelihood estimate (MLE) of $\theta$. The AIC function is
\[
\mbox{AIC}(k)= \!\!n_{k,L}-2(K^{L+1}-K^{k+1})(K-1),
\]
and the estimated order is $\hat{k} = \arg\min_{0\leqslant{k}\leqslant{L}}\mbox{AIC}(k)$.

Katz \cite{Katz81} applied the Bayesian information criterion (BIC) to the
problem of Markov chain order estimation. Similarly to AIC, BIC is defined as
\[
\mbox{BIC}(k)=\!\!n_{k,L}-(K^{L+1}-K^{k+1})(K-1)\ln{N},
\]
and the order estimate is $\hat{k}=\arg\min_{0\leqslant{k}\leqslant{L}}\mbox{BIC}(k)$. Though AIC is known to be  inconsistent and BIC consistent order estimate \cite{Csiszar00}, it was shown that BIC does not perform as well as AIC for small sample sizes \cite{Katz81,Csiszar00}.

The Peres-Shields estimator \cite{Pere05} uses the so-called fluctuation function
\[
\Delta^k_x(\upsilon)= \max_{a\in A}\left| N_x(\upsilon a)-\frac{N_x(\tau_k(\upsilon)a)}{N_x(\tau_k(\upsilon))}N_x(\upsilon) \right|,
\]
where $N_x(\upsilon)$ denotes the frequency of occurrence of the word $\upsilon$ of length
$l$ in $\{x_t\}_{t=1}^N$ and $\tau_k(\upsilon)$ denotes the $k$-suffix of $\upsilon$, i.e. $\tau_k(\upsilon)=\upsilon^l_{l-k+1}$.
The initial expression of the Peres-Shields estimator is rather complicated and Dalevi and Dubhashi \cite{Dale05} proposed a simpler estimator, still close to the original Peres-Shields estimator, given as
\[
\hat{k}=\arg\max_{k\geq0}(\Delta_{x}^{k}(\upsilon)/\Delta_{x}^{k+1}(\upsilon))
\]
and we use this estimator in the comparative study denoted as PS.

Menendez et al \cite{Mene01} start with the observation that the LR test can be expressed in terms of the Kullback-Liebler divergence, which belongs to the class of the so-called $\phi$-divergence measures. Then they generalize LR for orders $k$ and $k+1$ for any $\phi$-divergence given as
\[
S^{\phi}=\frac{2}{\phi''(1)}\sum_{a_{1},\dots,a_{k+1}}n_{a_1,\dots,a_k}\sum_{a_{k+1}}
\hat{p}(a_{k+1}|a_2,\dots,a_k) \phi\left(\frac{\hat{p}(a_{k+1}|a_1,\dots,a_k)}{\hat{p}(a_{k+1}|a_2,\dots,a_k)}\right),
\]
where $\phi''$ is the second derivative of $\phi$. For $\phi(x)=x \log{x}-x+1$ we get the standard LR in terms of Kullback-Liebler divergence. Menendez et al \cite{Mene11} suggest using $\phi(x)=(\lambda(\lambda+1))^{-1}(x^{\lambda+1}-x+\lambda(1-x))$ for $\lambda=2/3$, and they repeat LR test for increasing $k$ order until no rejection is obtained, where $S^{\phi}$ follows the Chi-squared distribution with $(K^{k+1}-K^k)(K-1)$ degrees of freedom. We adopt this form of the test in the comparative study and denote it Sf.

\section{Monte Carlo Simulations}
\label{sec:Simulations}

We compare the CMI-testing to the approximate CMI bias estimate of (\ref{eq:bias}), as well as other known criteria for the estimation of the Markov
chain order $L$, and for this we use Monte Carlo simulations for varying parameters $L$, $K$ and $N$. For each parameter setting, we use $100$
realizations and for the CMI-testing $M=1000$ randomized sequences for each realization, and for all estimation methods the order is sought in the
range $m=1,\ldots,L+1$. In the first simulation setup, Markov chains are derived by randomly set transition probability matrices of given order $L$,
while in the second simulation setup Markov chains are derived by transition matrices of given order $L$, estimated on two DNA sequence of genes and
intergenic regions. The results on the latter setting will give us the grounds for interpreting the results from the estimation of the Markov chain
order on the DNA sequences in Sec.~\ref{sec:DNA}.

\subsection{Randomly selected transition probabilities}

\begin{figure}[htb]
\centerline{\hbox{\includegraphics[width=7cm]{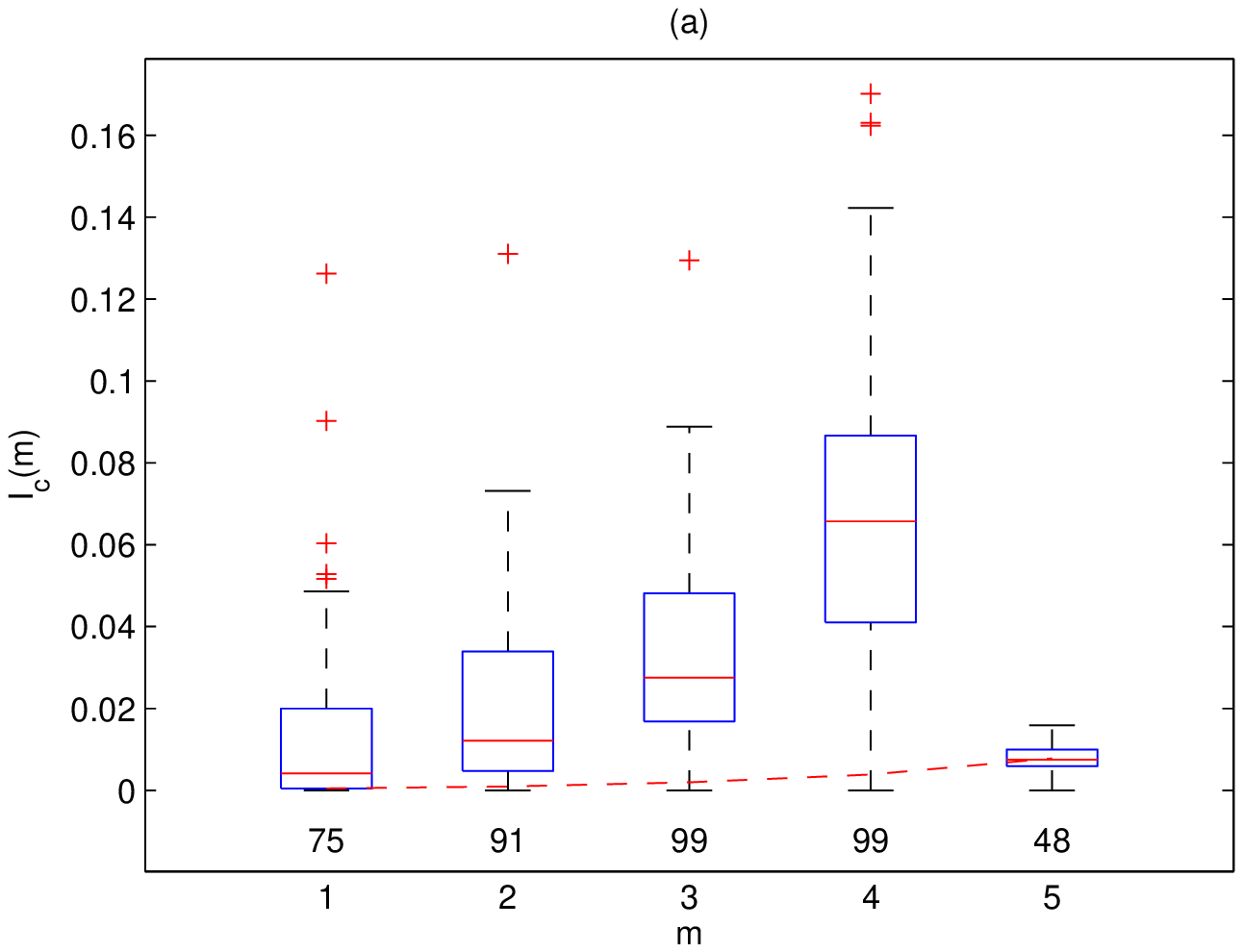}
\includegraphics[width=7cm]{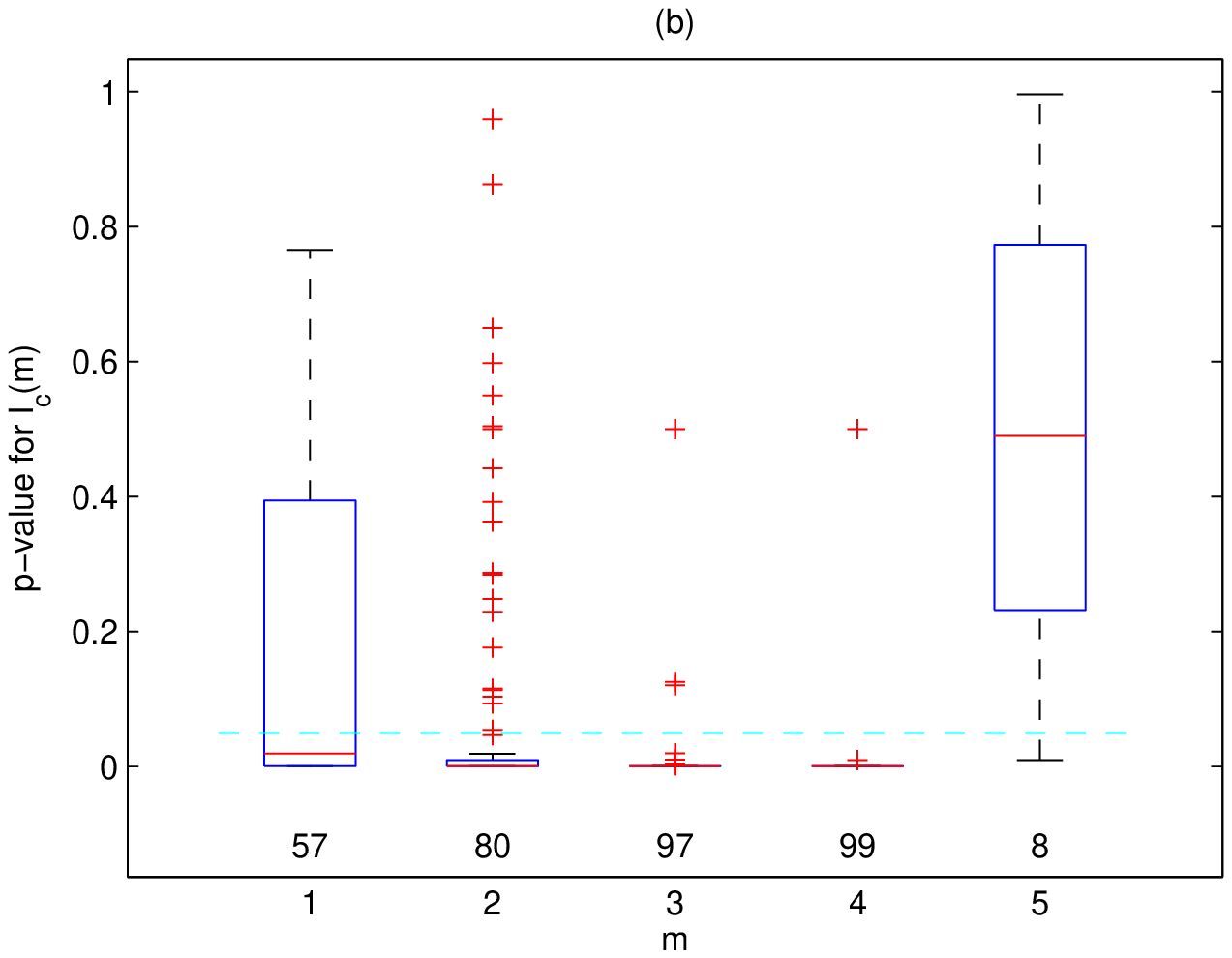}}}
\caption{Distribution of $\hat{I}_c(m)$ in (a) and $p$-values in (b) against the order $m$ presented as boxplots from 100 realizations of the Markov
chain as in Figure~\ref{fig:example}. The number below each boxplot is for the realizations for which $\hat{I}_c(m)$ is over the approximate
significance bias (displayed by a grey (cyan online) dashed line) in (a), and the $p$-value of the randomization significance test is below the
significance level $\alpha=0.05$ (displayed by a dashed line) in (b).} \label{fig:K2L4N1000}
\end{figure}

First, we confirm the results about the illustrative example of Figure~\ref{fig:example} using 100 realizations. As $m$ increases from 1 to $L$,
$\hat{I}_c(m)$ increases and lies over the approximate bias for $I_c(m)=0$ defined in (\ref{eq:bias}), as shown by the boxplots in
Figure~\ref{fig:K2L4N1000}a and indicated by the number of cases $\hat{I}_c(m)$ exceeding the limit of the bias approximation for each $m$.
However, when $m=L+1$, $\hat{I}_c(m)$ falls below this approximated bias for only about half of the 100 realizations, indicating that the approximate
bias cannot establish the significance of $I_c(m)$. On the other hand, significance is well-established by the proposed CMI-testing, and the
transition from significant to insignificant $I_c(m)$ at $m=L$ can be safely detected. As shown in Figure~\ref{fig:K2L4N1000}b, for all but one
realization $\mbox{H}_0: I_c(L)=0$ is rejected at the significance level $\alpha=0.05$, and only for 8 realizations $\mbox{H}_0: I_c(L+1)=0$ is
rejected at the same $\alpha$. We note that for $m=1$ the power of $\hat{I}_c(m)$ is very low (rejection is obtained for only 57 cases), and improves
as $m$ increases towards $L$. The first reason is that in general the randomization test is conservative \cite{Kugiumtzis08a}, e.g. note that
$\hat{I}_c(1)$ is over the significance bias limit far more often than the significance limit drawn by the randomized sequences. The second and most
important reason is specific to the simulation setup. The random selection of the transition matrix determines on average even dependence of one
symbol in the chain to the $L$ preceding symbols. Thus the knowledge of $x_{t-1}$ contributes partially (by a factor of about 1/4 for our example
with $L=4$) to the total information about $x_t$, and therefore $\hat{I}(x_t;x_{t-1})=\hat{I}_c(1)$ is small and can often be at the border of being
statistically significant. The bias of $\hat{I}_c(1)$ here is $1/(2N)$ (see (\ref{eq:bias}) for $K=2$) and thus very close to zero, so that the
distribution of $\hat{I}_c^{*i}(1)$ from the randomized sequences is asymmetric and more likely to be broader to the right than for a larger bias.
The information from $x_{t-2}$ about $x_t$ is at the same level as for $x_{t-1}$ when accounting for $x_{t-1}$, but now the bias of $\hat{I}_c(2)$ is
doubled and the distribution for the randomized sequences is more symmetric and less broader to the right, so that $\hat{I}_c(2)$ may be at the right
tail of the null distribution more often. This argument explains the increase of the percentage of rejections as $m$ increases from one to $L$.

We compare the CMI-testing, and refer to it simply as CMI, to four known criteria for the estimation of $L$: the Akaike's information criterion (AIC)
\cite{Tong75,Katz81,Guttorp95}, the Bayesian information criterion (BIC) \cite{Schwa78,Katz81,Csiszar00,Dalevi06}, the criterion of Dalevi and Dubashi which is based on the Peres and Shield's estimator (PS) \cite{Pere05,Dale05}, and the criterion of Menendez et al (Sf)
\cite{Mene06,Mene11}. Table~\ref{tab:randPK2} presents the frequency of estimating correctly $L$ for Markov chains of the first simulation setup,
$K=2$, $N=500,1000$ and $N=6000$.
\begin{table}
\caption{Number of times the correct order $L$ is estimated in 100 realizations by the criteria CMI, AIC, BIC, PS and Sf. The 100 symbol sequences of
length $N$ (being 500, 1000 and 6000 as indicated in the first row) are generated by a Markov chain of $K=2$ symbols with a randomly
selected transition probability matrix of order $L$, for $L$ varying as given in the second row. The best success rate for each $L$ is highlighted.}
\centering
\bigskip
\setlength{\tabcolsep}{0.3mm}
\begin{tabular}{|c|r|r|r|r|r|r|r|r|r|r|r|r|r|} \hline
{\em criterion} & \multicolumn{5}{c|}{$N=500$} & \multicolumn{4}{c|}{$N=1000$} & \multicolumn{4}{c|}{$N=6000$}\\ \hline & $L=2$ & $L=3$ & $L=4$ &
$L=5$ & $L=7$ & $L=2$ & $L=4$ & $L=6$ & $L=8$ & $L=2$ & $L=4$ & $L=6$ & $L=8$ \\ \hline
CMI & 81 &\textbf{92} & \textbf{95} &\textbf{94} & \textbf{73} & 87 & \textbf{91} & \textbf{98} & \textbf{91} & 93 & \textbf{98} & 95 & \textbf{97}\\
AIC & 66 & 62 & 52 & 38 & 2 & 75 & 68 & 33 & 2 & 77 & 80 &60 &36\\
BIC & 69 & 56 & 35 & 1 & 0 & 77 & 59 & 1 & 0 & 88 & 80 & 59 & 0\\
PS & 78 & 72 & 73 & 63 & 37 & 88 & 77 & 71 & 47 & \textbf{97} & \textbf{98} & \textbf{99} & 96\\
Sf & \textbf{86} & 60 & 46 & 28 & 0 & \textbf{92} & 42 & 29 & 0 & \textbf{97} & 3 & 1 & 40\\  \hline
\end{tabular}
\label{tab:randPK2}
\end{table}
For $L=2$, Sf scores highest with CMI being close behind, but for larger $L$ the success rate of Sf decreases steadily while CMI estimates the
correct order almost always, scoring much higher than all the other criteria. For the largest order $L=7$ examined for $N=500$, the success rate of
CMI decreases, probably due to insufficient data size for such a large order, but the other criteria fail completely to estimate this order and only
PS manages it for 37 of the realizations. All methods improve their performance when the sequence length increases to $N=1000$ but at about the same
degree so that the main differences persist. For larger $L$ CMI maintains the highest success rate at a level over 90\%, even for $L=8$, while the
other criteria fail, with Sf dropping again to the zero level for $L=8$. AIC and BIC follow a similar decreasing success rate with $L$ down to the
zero level with BIC being worse, and PS attains higher success rates for larger $L$ but still much lower than for CMI. When the chain
length further increases ($N=6000$), PS improves and performs as well as CMI.

The estimation of $L$ is more data demanding when there are more symbols $K$. As shown for $K=4$ in Table~\ref{tab:randPK4}, for
$N=500$, though CMI, PS and Sf succeed to identify the correct order for $L=2,3$ (with CMI scoring highest), all but PS fail for $L>3$ with the score
of PS falling slowest. For larger $N$ ($N=1000$ and $N=6000$) the failure of the criteria occurs for larger $L$.
\begin{table}
\caption{Same as for Table~\ref{tab:randPK2}, but for $K=4$.} \centering
\bigskip
\setlength{\tabcolsep}{0.3mm}
\begin{tabular}{|c|r|r|r|r|r|r|r|r|r|r|r|r|r|} \hline
{\em criterion} & \multicolumn{4}{c|}{$N=500$} & \multicolumn{5}{c|}{$N=1000$} & \multicolumn{4}{c|}{$N=6000$} \\ \hline & $L=2$ & $L=3$ & $L=4$ &
$L=5$ & $L=2$ & $L=3$ & $L=4$ & $L=5$ & $L=6$ & $L=2$ & $L=4$ & $L=5$ & $L=6$ \\ \hline
CMI & \textbf{100} & \textbf{100} & 18 & 1 & 98 & \textbf{100} & \textbf{96} & 2 & 3 & 96 & \textbf{100} & \textbf{100} & 5 \\
AIC & 0 & 0 & 0 & 0 & 2 & 0 & 0 & 0 & 0 & 37 & 0 & 0 & 0 \\
BIC & 0 & 0 &0 & 0 & 0 & 0 & 0 & 0 & 0 & 32 & 0 & 0 & 0 \\
PS & 96 & 81 & \textbf{46} & \textbf{23} & \textbf{100} & 89 & 46 & \textbf{21} & \textbf{14} & \textbf{100} & \textbf{100} & 61 & \textbf{19} \\
Sf & \textbf{100} & 72 & 0 & 0 & \textbf{100} & \textbf{100} & 0 & 0 & 0 & \textbf{100} & \textbf{100} & 62 & 0 \\  \hline
\end{tabular}
\label{tab:randPK4}
\end{table}
AIC and BIC fail completely and only for $L=2$ and $N=6000$ they have a success rate at about one third. Again CMI keeps the high success rate as $L$
increases until it collapses due to lack of sufficient data, but so do the other criteria already for smaller $L$. For example, for $L=3$ and
$N=1000$ both CMI and Sf score highest, but for $L=4$ CMI still scores very high while Sf has dropped to zero score. Generally, Sf has the tendency
to underestimate the order for larger $L$. Specifically, for the above simulation when $L=4$, Sf estimates $m=1,m=2,m=3$ at the rates
$52\%,32\%,16\%$ respectively. For larger $L$, PS tends to maintain some positive success rate when all other criteria fail completely (almost
completely for CMI).

The results of the simulation setup of randomly selected transition matrices showed that CMI overall outperforms the other criteria, whereas PS
scores well for large $L$ (at cases even higher than CMI), and Sf is best for very small $L$ but scores poorly for larger $L$. AIC
and BIC perform well for small number of symbols $K$, but their requirement for data size increases faster with $K$ than for the other criteria. BIC
tends to perform better than AIC for very small $L$, but this situation is reversed when $L$ increases. PS and CMI keep the highest rate for larger
orders over all settings. However, CMI stays ahead when the length of symbol sequences is smaller, while the success rate of PS seems to fall slower
when the order becomes larger.

\subsection{Transition probabilities estimated on DNA}

DNA consists of four nucleotides, the two purines, adenine (A) and guanine (G), and the two pyrimidines, cytosine (C) and thymine (T), so DNA
sequence can be considered as a symbolic sequence on the symbols A,C,G,T. In our analysis we use a large segment of the Chromosome 1 of the plant
\emph{Arabidopsis} \emph{thaliana}. We use two sequences, one joining together the genes, which contain non-coding regions, called introns, in
between the coding regions, called exons, and another sequence joining together the intergenic regions which have non-coding character. The sequences
used here are segments of the long sequences used in \cite{Kugiumtzis04c}.

For the second simulation setup we form the Markov chains from transition matrices of given order $L$ estimated on the two DNA sequences of genes and
intergenic regions, each of length $N=6000$. We make the simulations for $K=4$ symbols (A, C, G, T) and $K=2$ symbols (purines, pyrimidines).

The probability transition matrices of any order $L$ estimated on the DNA sequences give more complicated structures of the Markov chains and make
the estimation of $L$ harder than when they are randomly selected. As shown in Table~\ref{tab:simDNAcoding} for the gene sequence, all criteria score
lower than for the respective orders $L$ of the first simulation setup even for very small $L$, though we use quite large sequences ($N=6000$).
\begin{table}
\caption{Same format of results as for Table~\ref{tab:randPK2}, but for transition matrices of given order $L$ estimated from a DNA sequence of genes
of length $N=6000$ in the form of purines and pyrimidines ($K=2$) and all the four nucleotides ($K=4$).} \centering
\bigskip
\setlength{\tabcolsep}{2mm}
\begin{tabular}{|c|r|r|r|r|r|r|r|r|} \hline
{\em criterion} & \multicolumn{4}{c|}{$K=2$} & \multicolumn{4}{c|}{$K=4$} \\ \hline & $L=2$ & $L=3$ & $L=4$ & $L=5$ & $L=2$ & $L=3$ & $L=4$ & $L=5$\\
\hline
CMI & 50 & 65 & \textbf{15} & 8 & 93 & \textbf{66} & \textbf{35} & 0\\
AIC & \textbf{63}  &\textbf{82} & \textbf{15} & 2 & 87 & 0 & 0 & 0\\
BIC & 4 & 0 & 0 & 0 & 0 & 0 & 0 & 0\\
PS & 56 & 58 & 13 & \textbf{9} & 73 & 42 & 14 & \textbf{8}\\
Sf & 50 & 20 & 1 & 0 & \textbf{99} & 19 & 0 & 0\\  \hline
\end{tabular}
\label{tab:simDNAcoding}
\end{table}
CMI is generally best for $K=4$, e.g. for $L=3$ CMI estimates the correct order for 2/3 of the realizations with PS being second best estimating
correctly for 42 realizations, Sf for only 19, and AIC and BIC for none. However, AIC performs better than the other criteria when $K=2$, being best
for $L=2$, $L=3$ and $L=4$ followed by PS and CMI (and Sf for $L=2$). For both $K=2$ and $K=4$, as $L$ increases the percentage of success rate falls
sharply for the other criteria but more regularly for CMI and PS.

For the intergenic regions the results are somehow better for all criteria. As shown in Table~\ref{tab:simDNAnoncoding}, for $K=2$ AIC scores highest
for $L=2$ and $L=4$, with CMI following very closely (and Sf only for $L=2$).
\begin{table}
\caption{Same as for Table~\ref{tab:simDNAcoding}, but for the DNA sequence of intergenic regions.} \centering
\bigskip
\setlength{\tabcolsep}{1.5mm}
\begin{tabular}{|c|r|r|r|r|r|r|r|r|r|} \hline
{\em criterion} & \multicolumn{5}{c|}{$K=2$} & \multicolumn{4}{c|}{$K=4$} \\ \hline & $L=2$ & $L=4$ & $L=5$ & $L=6$ & $L=7$ & $L=2$ & $L=3$ & $L=4$ &
$L=5$ \\ \hline
CMI & 81 & 83 & \textbf{50} & \textbf{44} & \textbf{16} & 94 & \textbf{66} & \textbf{45} & 8\\
AIC & \textbf{85} & \textbf{88} & 27 & 1 & 0 & 79  & 0 & 0 & 0 \\
BIC & 42 & 0 & 0 & 0 & 0 & 0 & 0 & 0 & 0\\
PS & 81 & 43 & 27 & 28 & 14 & 92 & 28 & 10 & \textbf{21} \\
Sf & \textbf{85} & 9 & 1 & 0 & 0 & \textbf{98} & 23 & 1 & 0 \\  \hline
\end{tabular}
\label{tab:simDNAnoncoding}
\end{table}
As $L$ increases the percentage of success rate falls sharply for the other criteria but more regularly for CMI and PS, while CMI scores highest
maintaining a success rate at about 50\% for $L=5$ and $L=6$. The same rapidly decreasing success rate for $L>2$ holds for $K=4$ and for all but CMI
criteria. Nevertheless, CMI fails also to estimate the correct order for $L>6$ and $L>4$ when $K=2$ and $K=4$, respectively. The latter indicates the
limit of orders that can be estimated with CMI for $N=6000$, so that if the real DNA sequence has larger order (or even infinite) this could not be
estimated by CMI with such limited sequence.Generally, AIC outperforms the other criteria for smaller orders and fewer symbols.
However, CMI and PS score highest for larger $L$, while CMI performs better than PS for larger $K$.

\section{Application on DNA sequences}
\label{sec:DNA}
In recent years, much of the statistical analysis of DNA sequences is focused on the estimation of properties of coding and non-coding regions as
well as on the discrimination of these regions. There has been evidence that there is a different structure in coding and non-coding sequences and
that the non-coding sequences tend to have long range correlation, whereas the correlation in coding sequences exhibits exponential decay
\cite{Peng92,Buldyrev98,Almirantis99}. Here we use intergenic and gene sequences. The latter is a mixture of coding regions (exons) and non-coding
regions (introns), and therefore we expect to have also long correlation due to the non-coding regions in it, but it should be less than the
correlation in the intergenic regions consisting only of non-coding parts. Thus both DNA sequences cannot be considered as Markov chains, at least
not of a moderate order, and the estimation of the order $L$ should increase with the data size.

\begin{figure}[htb]
\centerline{\hbox{\includegraphics[width=7cm]{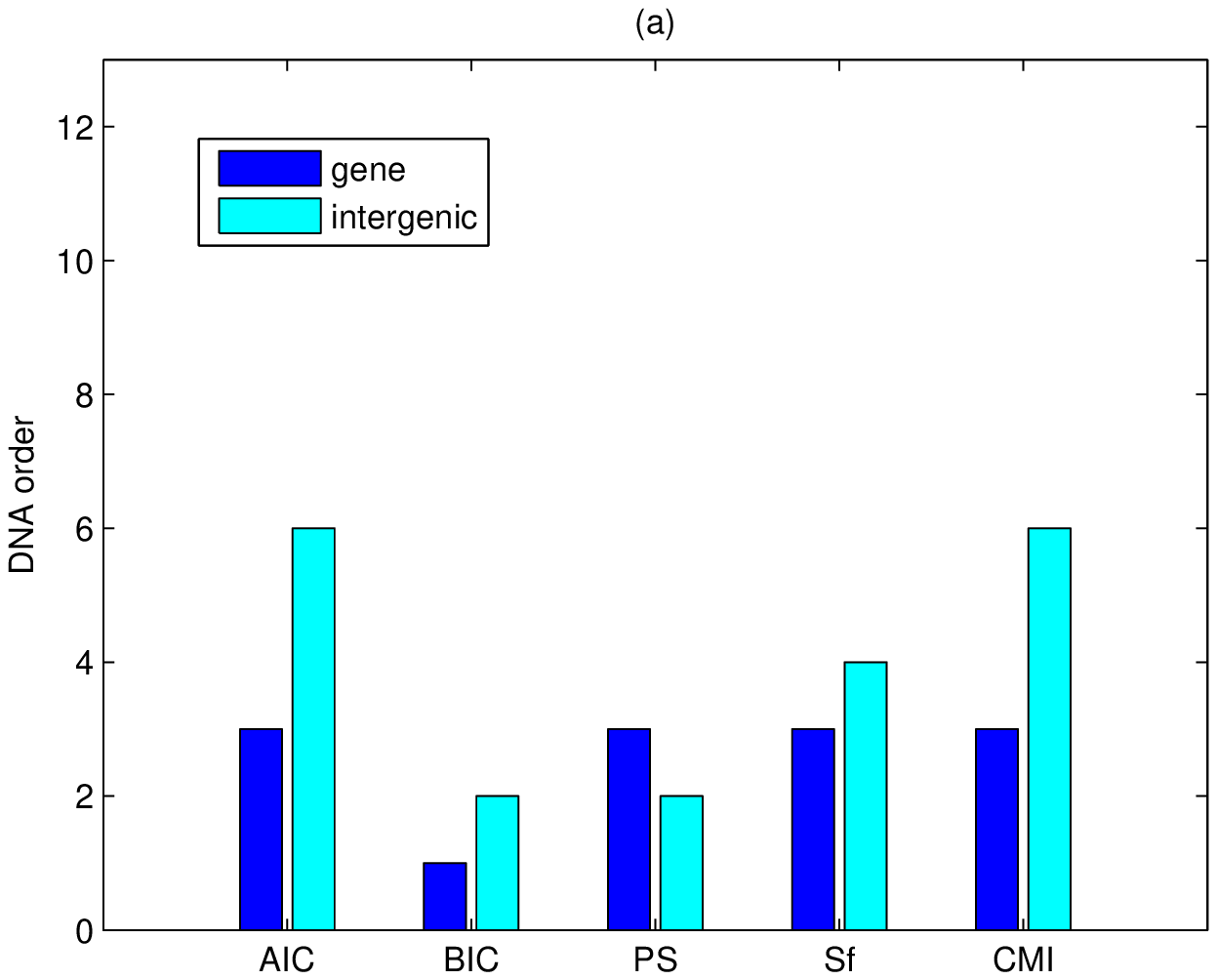}
\includegraphics[width=7cm]{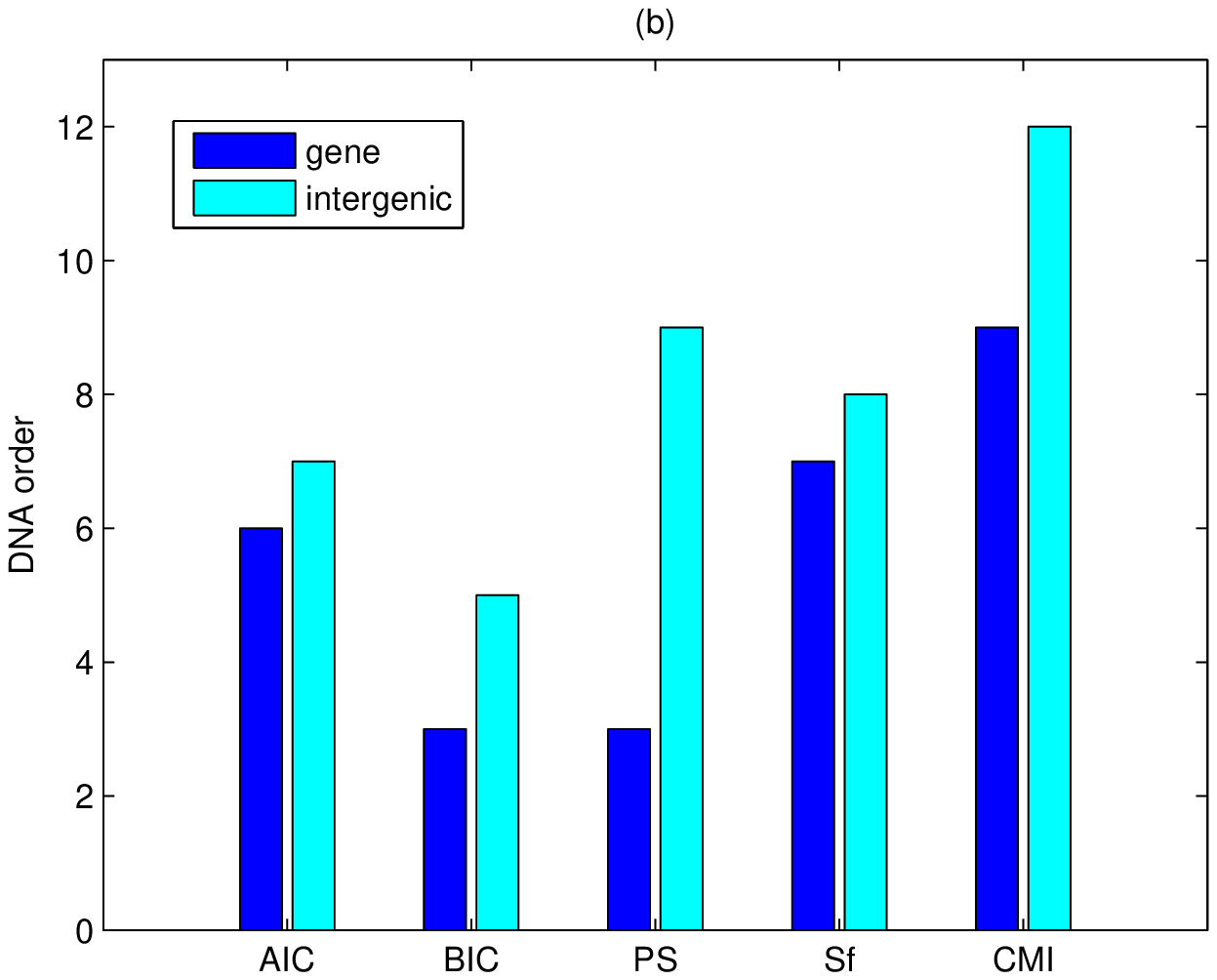}}}
\caption{The estimated order $L$ of a Markov chain on Chromosome 1 of plant A\emph{rabidopsis} \emph{thaliana} (genes and intergenic regions) by the
CMI-testing and the other criteria. The computations are made for number of symbols $K=2$ and length $N=10000$ in (a) and $N=100000$ in (b).}
\label{fig:estimationorderDNA}
\end{figure}

We estimate the order $L$ of a hypothesized Markov Chain on Chromosome 1 of plant A\emph{rabidopsis} \emph{thaliana} by the CMI-testing and the other
criteria. We make the computations for both genes and intergenic regions of length $N=10000$ and $N=100000$ and for $K=2$ (purines, pyrimidines), and
the estimated orders from all criteria are shown in Figure~\ref{fig:estimationorderDNA}.
First we note that all criteria tend to estimate larger order as $N$ increases, and for the same $N$ they find larger order for the intergenic
sequence, both features being in agreement with the discussion above. CMI establishes best these two features. The difference in the order of genes
and intergenic regions holds for both $N$ and their orders increase the most from 3 and 6 for $N=10000$ to 9 and 12 for $N=100000$, respectively. AIC
estimates the same orders as CMI for $N=10000$, but for $N=100000$ only the estimated order for genes increases to 6 approaching the order for the
intergenic region staying at about the same level. The other three criteria estimate smaller orders than CMI and AIC for $N=10000$. Moreover, PS
gives for $N=10000$ the reverse pattern of the order for genes being 3 and for intergenic regions being 2, which changes to 3 and 9 for $N=100000$,
respectively. BIC and Sf give order estimates closer to the expected two features, but the order estimation is at a lower level than for CMI with Sf
giving larger orders than BIC. CMI is the most consistent to the hypothesis of long range correlation in the intergenic sequence, and at a lesser
degree to the gene sequence, as it provides the largest dependence of the order to the sequence length and maintains larger order for the intergenic
sequence.

\section{Discussion}
\label{sec:Discussion}
In this work we propose the use of the measure of conditional mutual information (CMI) for the estimation of the order of Markov chain, in an
analogous way the partial autocorrelation is used for the estimation of the order of an autoregressive model in time series \cite{Box94a}. Among
others, a main difference is that the significance limits for partial autocorrelation are defined parametrically (under mild conditions), while for
CMI only approximate limits have been reported. Our simulations on analytic limits for the bias of CMI, which we have worked out, showed that they
cannot provide accurate estimation of the Markov chain order $L$. Therefore we have built a scheme called CMI-testing, applying iteratively a
randomization significance test for CMI, and the estimation of $L$ is given by the largest order $m$ for which CMI is found statistically
significant. Thus CMI-testing does not implicate any maximum order, as for example the criteria of AIC and BIC.

We compared CMI-testing to a number of other known order selection criteria using Monte Carlo simulations of Markov chains of varying order $L$ and
number of symbols $K$, and for different sequence lengths $N$. Randomization tests tend to be more conservative for small data sizes, but we found
that CMI-testing could identify the correct $L$ even at small sequences, e.g. for $K=2$, $N=500$ and $L=5$ the success rate was 94\%. For larger $K$
and $L$, and for smaller $N$, the accuracy of the estimation worsened, but still compared to the other criteria it was generally the highest. For
small $L$, other criteria could score higher but CMI-testing always followed closely.

The simulations showed the appropriateness of CMI-testing in the settings of nontrivial structures in the symbol sequences, involving high Markov
chain order $L$. This was further confirmed by the simulations on Markov chains estimated on DNA sequences, but also when applied, along with other
criteria, to two real DNA sequences, one comprised of genes and the other of intergenic regions. Many reported works converge to that intergenic
regions (consisting solely of non-coding DNA) have long range correlations, and genes (containing coding and non-coding DNA) have a mixture of short
and long range correlations. As the estimation of CMI is computationally intensive, we made computations on DNA sequences up to the length
$N=100000$, for which CMI-testing gave the largest Markov chain orders 9 and 12 for the genes and intergenic sequences, respectively, being both
higher than the orders obtained by any of the other criteria. This confirms the ability of CMI-testing in identifying large orders, as confirmed also
in the simulations.

To the best of our knowledge, this is the first work using CMI for the estimation of Markov chain order, and it certainly bears further improvement.
For the randomization significance test we use randomly shuffled sequences irrespective of the order $L$, and we attribute to this lack of any
dependence in the surrogate sequences the observation that for orders larger than the correct order $L$ the original CMI is often at the lower tail
of the distribution of the CMI on the surrogates. One possible improvement is to adjust the resampled sequences to the tested order, e.g. to be
generated from Markov chains of order being one less than the tested order. However, the generation of such randomized sequences is not
straightforward and it would additionally add to the heavy computational cost in CMI-testing. The latter is a disadvantage of CMI-testing in problems
were computation time may be an issue or when the sequence length is very large as for DNA. A parametric significance test would be a solution, which
may come at the cost of reduced accuracy as, to the best of our knowledge, there is no exact analytic distribution of CMI. We currently work on this
issue developing approximations for the CMI distribution.

\bibliographystyle{elsarticle-harv}


\end{document}